\documentstyle[12pt]{article}
\begin{document}
\title{Light Monopoles, Electric Vortices and Instantons}
\author{
Juan Mateos Guilarte \\ Department of Physics. Facultad de Ciencias\\
Universidad de Salamanca. Salamanca 37008\\
SPAIN}
\maketitle
\begin{abstract}
A Quantum Field Theory for magnetic monopoles is described and its phase
structure fully
analysed.
\end{abstract}

\section{}
The strong coupling limit of $N=3D2$ supersymmetric gauge theory in four
dimensions has been studied in depth by Seiberg and Witten in a brilliant
series of papers \cite{s-w}. An important outcome is that the 'strongly
coupled' vacuum turns out to be a weakly coupled theory of light magnetic
monopoles.
Moreover, a version of electric-magnetic duality fitting
the Montonen-Olive conjecture \cite{m-o} proves to be true in the model.

In this letter we offer a brief description of the physics involved in
the Seiberg-Witten action. There are two kinds of quanta: light monopoles
and dual-photons. Monopole-monopole as well as dual-photon-monopole
interactions are due to the vertices coming from the non-quadratic terms of
the Lagrangian. This is the content of perturbation theory although=
 different
topological sectors show important non-perturbative effects in the model. We
find topological defects as the solutions of a minor modification,a sign,
of the monopole equations \cite{wit1}. The change yields non-trivial regular
solutions
in ${\bf R}^4$ which, on physical grounds, we find to be either electric
vortices or
instantons. As a consequence, the phase
structure of the theory is determined.

We start from the euclidean action
\begin{eqnarray*}
S&=3D& \int d^4x\{\frac{1}{4}F_{\mu\nu}F_{\mu\nu}+\frac{1}{2}(D_\mu\Psi_+)^+
D_\mu\Psi_+ +\frac{1}{4}(D_\mu\Psi_+)^+\Sigma_{\mu\nu}D_\nu\Psi_+\}\\
 &+& \int d^4x\{i\frac{\lambda_1}{2}\varepsilon_{\mu\nu\rho\sigma}
 F_{\rho\sigma} j_{\mu\nu}+\frac{\lambda_2^2}{8}j_{\mu\nu}j_{\mu\nu}\}.
 \hspace{6cm}(1)
\end{eqnarray*}
$\Psi_+$ is a right-handed Weyl spinor, the light monopole field. $A_\mu$
is a dual electromagnetic potential. $D_\mu\Psi_+=3D\partial_\mu \Psi_++ig
A_\mu \Psi_+$
is the covariant derivative with respect to the gauge group $U(1)_d$, the
dual of the electromagnetic $U(1)$. The magnetic charge,$g=3D\frac{4\pi}{e}$
is the coupling constant and the tensor field is $F_{\mu\nu}=3D\partial_\mu
A_\nu -\partial_\nu A_\mu$. Finally,=
 $\Sigma_{\mu\nu}=3D\frac{1}{2}[\gamma_\mu,
\gamma_\nu]$ where, $\gamma_\mu$ are the euclidean $\gamma$-matrices in=
 chiral
representation: if $\sigma_i$ are the Pauli matrices,
\[ \gamma_\mu =3Di\left( \begin{array}{cc} 0 & \sigma_\mu \\
                         {\overline{\sigma}}_\mu & 0 \end{array} \right)
        \hspace{.5cm}
        \begin{array}{lll} \sigma_\mu & =3D & (i1_2,\sigma_i) \\
                           {\overline{\sigma}}_\mu & =3D & (i1_2,-\sigma_i)
        \end{array} \hspace{.5cm}
        \gamma_s=3D \left( \begin{array}{cc} -I & 0 \\ 0 & I \end{array}=
 \right)
        \]
\[ \Psi (x)=3D\left( \begin{array}{c} \Psi_L (x) \\ \Psi_R (x) \end{array}
             \right) \hspace{.5cm}
  \Psi_{\pm} (x)=3D\frac{1\pm \gamma_5}{2}\Psi (x)   \]
and $j_{\mu\nu} (x)=3D\frac{1}{2} \Psi_+^\dagger (x) \Sigma_{\mu\nu}
\Psi_+ (x).$

The unusual kinetic terms in the spinor fields are possible because in=
 euclidean
space there is $SO(4)$ symmetry: there are no linear scalars in the=
 derivatives
for Weyl spinors, (see \cite{ram}). These are due to the duality of the
monopoles
with respect to the scalar fields of the parent theory, the $N=3D2$=
 supersymmetric
Yang-Mills-Higgs theory \cite{ow}. Despite the existence of terms of the=
 form
$j_{\mu\nu}j_{\mu\nu}$
the theory is renormalizable; the dimension of $\Psi_+$ is 1, a non-crucial=
 fact
because we take the model as physically relevant in the strong $e$, weak=
 $g$,
limit.

The $j-F$ interaction term in (1) tells us that physically $j_{\mu\nu}$ is
due to
the electric dipolar momentum of the monopole. We deal with a spinor
representation
of the Lorentz group twisted by the $U(1)$ group of electro-magnetic duality
transformations carrying a property that we call $m$-spin. The $m$-spin, or
$m$-helicity in the massless case, yields an intrinsic electric dipolar=
 momentum
in the magnetic monopoles and labels the representations of this strange=
 version
 of the Lorentz group which presumably comes from some twisting of the=
 original
$N=3D2$ $SYMH$ theory.

In this model, on the other hand, all the fields are in the adjoint
representation
of the $SU(2)$ gauge group. Thus, the spin of the monopole is zero meanwhile
the dyonic excitations have an integer spin: the monopole field, despite its
spinorial nature, must be quantized according to Bose statistics.

Perturbation theory tells us that there are
two kinds of quanta:
\begin{enumerate}
\item Light Monopoles. The momentum space propagator is:
\[ \Delta _E^M (k)=3D\frac{k_\mu=
 (\delta_{\mu\nu}+{\bar{\sigma}}_{\mu\nu})k_\nu}
{k^4} \hspace{.5cm},\hspace{.5cm}
  \begin{array}{lcl} {\bar{\sigma}}_{ij}& =3D & i\varepsilon_{ijk}\sigma_k=
 \\
                     {\bar{\sigma}}_{i4}&=3D& i\sigma_i=3D-{\bar{\sigma}}_{4=
i}
  \end{array}. \]

\item Dual-photons. The euclidean propagator is:
\[ \Delta_{E_{\mu\nu}}^d (k)=3D\frac{1}{k^2} (\delta_{\mu\nu}+(\alpha-1)
\frac{k_\mu k_\nu}{k^2}) .\]
We also read from the non-quadratic terms of the Lagrangian two kinds of
vertices:
\begin{itemize}
\item{a.} A quartic monopole self-interaction:
\[ V_{MM}=3D\lambda_2^2 {\bar{\sigma}}_{\mu\nu} {\bar{\sigma}}_{\mu\nu} \]
\item{b.} A dual-photon/monopole vertex:
\[ V_{MD}(k)=3Dgk_\mu +\lambda_1\varepsilon_{\mu\nu\rho\sigma}
  {\bar{\sigma}}_{\nu\rho}k_\sigma \]
\end{itemize}
Physical features: besides renormalizability, at least at the same level as
scalar QED, perturbation theory is conformally invariant and the electric
dipolar
momentum contributes both to the MD and the MM vertices; not only the=
 magnetic
charge $g$ and the $\lambda_2$ coupling are important.

The amplitude for monopole-monopole scattering at the lowest order in
perturbation theory and
$\alpha=3D1$, the Feynman gauge, is:
\begin{eqnarray*} T_{MM}(k_2^\prime,k_2;k_1^\prime,&k_1&)=3D
 U_{1/2}^\dagger (k_2^\prime)(gk_\mu+\lambda_1\varepsilon_{\mu\nu\rho\sigma}
   {\bar{\sigma}}_{\nu\rho}k_\sigma)U_{1/2}(k_2) \\
& &\frac{1}{k^2}\cdot U_{1/2}^\dagger (k_1^\prime)(gk_\mu+\lambda_1
 \varepsilon_{\mu\alpha\beta\gamma}
 {\bar{\sigma}}_{\alpha\beta}k_\gamma)U_{1/2}(k_1) \\
& & \delta^{(4)} (k_1+k_2-k_1^\prime-k_2^\prime)+ \lambda_2^2 U_{1/2}^+
(k_2^\prime)
{\bar{\sigma}}_{\mu\nu}U_{1/2}(k_2) \\
& & U_{1/2}^+(k_1^\prime){\bar{\sigma}}_{\mu\nu}U_{1/2}(k_1) \delta^{(4)}
(k_1+k_2-k_1^\prime-
k_2^\prime)
\end{eqnarray*}
if $k_\mu=3Dk_{2\mu}^\prime-k_{2\mu}=3Dk_{1\mu}^\prime-k_{1\mu}$ and=
 $U_{1/2} (k)$
are the plane wave spinors:
\[ U_{1/2} (k)=3D\left( \begin{array}{cc} 1 \\ \frac{k_1+ik_2}{|\vec{k}|+
                                   k_3} \end{array} \right) .\]
\end{enumerate}

\section{}
To study the non-perturbative regime, it is convenient to implement a=
 Bogomolny
splitting \cite{bog}: defining the self-dual part of $F_{\mu\nu}$ as
$ F_{\mu\nu}^+ =3D\frac{1}{2} (F_{\mu\nu}+\varepsilon_{\mu\nu\rho\sigma}
    F_{\rho\sigma}) $

we write:
\[\begin{array}{l} S=3D \int d^4x \{ \frac{1}{2} (F_{\mu\nu}^+-
   \frac{i}{2} \lambda_{2} j_{\mu\nu})^2+\frac{1}{4} (D_\mu \Psi_+)^+=
 \gamma_\mu
   \gamma_\nu D_\nu\Psi_+\} \\
   +\int d^4x \{\frac{i\lambda_2}{2} F_{\mu\nu}^+j_{\mu\nu}+\frac{1}{2}
   (D_\mu\Psi_+)^+\Sigma_{\mu\nu}D_\nu\Psi_++\frac{i\lambda_1}{4}
   \varepsilon_{\mu\nu\rho\sigma}F_{\rho\sigma}\Psi_+^\dagger=
 \Sigma_{\mu\nu}
   \Psi_+ \} \\
   -\int d^4x \{ \varepsilon_{\mu\nu\rho\sigma}F_{\mu\nu}F_{\rho\sigma} \}
   \hspace{8.3cm}(2)              \end{array} \]
If $\lambda_2=3Dg=3D-\lambda_1$, a critical point analogous to that=
 occurring
between type I and type II superconductors, (2) becomes
\begin{eqnarray*} S&=3D& \int d^4x \{ \frac{1}{2} (F_{\mu\nu}^+-\frac{i}{2}
           gj_{\mu\nu})^2+\frac{1}{4} (D_\mu\Psi_+)^\dagger \gamma_\mu
           \gamma_\nu D_\nu\Psi_+\} \\
           & -& \int d^4x \{\varepsilon_{\mu\nu\rho\sigma}F_{\mu\nu}
           F_{\rho\sigma} \} \end{eqnarray*}
and re-scaling $x\to \frac{1}{g}x$ we see that at the critical point $S$ is
almost $\frac{1}{g^2}$ times the Seiberg-Witten action plus a topological
term looking similar to the Pontryagin or second Chern number. There is a
different relative sign
in the first squared term; we choose the sign of the Pauli momentum=
 interaction,
the term $j-F$ term, in (1) by assigning the negative $m$-spin projection
to the monopole quanta. This amounts to a different choice of orientation in
${\bf R}^4$ with respect to the convention assumed in the  Seiberg-Witten=
 theory
due to the specific choice of the twisting in the $N=3D2$ SYMH parent=
 theory.

Solutions of the first order system,
\[ \begin{array}{lr}
F_{\mu\nu}^+=3D\frac{1}{2}\Psi_+^\dagger \Sigma_{\mu\nu}\Psi_+ &=
 \hspace{5.6cm}
                                     (3.a) \\
\gamma_\mu D_\mu\Psi_+ =3D 0 & \hspace{6.6cm} (3.b)
\end{array} \]
are absolute minima of the euclidean action and therefore play an important
r\^ole in the system.
In order to find them,
it is convenient to split (3.a-b) into components: (3.a) reads
\[ \begin{array}{lclr}
F_{12}^+=3D & -\frac{1}{2}(\phi_1^* \phi_1-\phi_2^* \phi_2) &=3DF_{34}^+ &
                          \hspace{4.3cm}(4.a) \\
F_{23}^+=3D & -\frac{1}{2}(\phi_1^* \phi_2+\phi_2^* \phi_1) &=3DF_{14}^+ &
                          \hspace{4.3cm}(4.b) \\
F_{13}^+=3D & \frac{i}{2}(\phi_1^* \phi_2-\phi_2^* \phi_1) &=3DF_{24}^+ &
                          \hspace{4.3cm}(4.c)
 \end{array} \]
 where $\Psi_R=3D\left( \begin{array}{c} \phi_1 \\ \phi_2 \end{array}=
 \right) $.
  Also, we have for (3.b)
 \[ \begin{array}{lr}
 (D_3+iD_4)\phi_1+i(D_1-iD_2)\phi_2=3D0 & \hspace{4.4cm}(5.a) \\
 (D_1+iD_2)\phi_1-i(D_3-iD_4)\phi_2=3D0 & \hspace{4.4cm}(5.b) \end{array} \]
 Multiplying (5.a) by $(D_1+iD_2)$, (5.b) by $(D_3+iD_4)$, subtracting,=
 using
 (4.b-c) and integrating over all ${\bf R}^4$ we obtain
 \[ \int d^4x \{ |\phi_1 |^2|\phi_2|^2+|D_{z_1}\phi_2|^2+|D_{z_2}\phi_2|^2=
 \}
 =3D0 \hspace{3.5cm}(6) \]
 $D_{z_1}=3DD_1-iD_2,\; D_{z_2}=3DD_3-iD_4$, after a partial integration. We=
 are
then
left with two possibilities:
\begin{itemize}
\item{A.} $\phi_2=3D0$. Solutions with $m$-spin 1/2 in the $x_3$-direction
$ \frac{1}{2}\sigma_3 \left( \begin{array}{c} \phi_1 \\ 0 \end{array}\right)
   =3D\frac{1}{2}\left( \begin{array}{c} \phi_1 \\ 0 \end{array} \right) $
\item{B.} $\phi_1=3D0$ and $D_{z_1}\phi_2+D_{z_2}\phi_2=3D0$. Solutions with
$m$-spin projection $-1/2$:
\[ \frac{1}{2}\sigma_3 \left( \begin{array}{c} 0 \\ \phi_2 \end{array}=
 \right)
   =3D-\frac{1}{2}\left( \begin{array}{c} 0 \\ \phi_2 \end{array} \right).=
 \]
  \end{itemize}
   Formula (6) tells us that type A and B exclude each other: even for
solitons,  only one of the two
polarizations is possible.
\section{}
Searching for explicit solutions we make the following ansatz adapted to=
 type
A:
\[ \phi_1(x)=3D\phi_1(x_1,x_2)\hspace{3cm} \phi_2=3DA_3=3DA_4=3D0 \]
\[ A_1(x)=3DA_1(x_1,x_2)\; ; \; A_2(x)=3DA_2(x_1,x_2) \]
The only non-zero (4)-(5) equations reduce to
\[ \begin{array}{l} F_{12}(x_1,x_2)=3D-|\phi_1(x_1,x_2)|^2 \\=7F
       (D_1+iD_2)\phi_1(x_1,x_2)=3D0 \end{array} \hspace{5.9cm}(7)\]
It is well known that system (7) is tantamount to the Liouville equation in
${\bf R}^2$,
\cite{wit2}, and the general solution such that $\lim\limits_{r_1 \to\infty}
\phi (x_1,x_2)=3D0$, where $r_1^2=3Dx_1^2+x_2^2$, guaranteeing finite energy
density is, see \cite{jack},
\[ \phi_1^{[k]}(z_1)=3D\frac{2 f'=
 (z_1)V^2(z_1)}{|V(z_1)|^2+|f(z_1)V(z_1)|^2}
\hspace{1cm}z_1=3Dx_1+ix_2 \hspace{2.3cm} (8) \]
\[ f(z)=3Df_0+\sum_{i=3D1}^k \frac{c_i}{z-z^{(i)}}\hspace{.5cm}=
 V(z)=3D\prod_{(i=3D1)}^k
(z-z^{(i)}) \]
The solutions $\phi_1^{[k]}$ have infinite action,
\[ S[\phi_1^{[k]}]=3D\frac{2\pi k}{g^2}\lim_{L,T\to\infty}
\int_{-\frac{L}{2}}^{-\frac{L}{2}}dx_3\int_{\frac{T}{2}}^{-\frac{T}{2}}=
 dx_4\]
although the electric flux is finite:
$\Phi_E[\phi_1^{[k]}]=3D\int d^2x\, F_{12}=3D\frac{2\pi}{g}k .$

Observe that $k$ is positive and the flux is located around $z_1^{(i)}$,
the zeroes of $\phi_1(z_1)$. It spreads out, however, with $|c_i|$, the=
 length
scale of the solution, which is a free parameter due to the scale invariance
of the
theory. There is also freedom in choosing $\mbox{arg} \,c_i$ because the
$U(1)_d$ symmetry and the moduli space of solutions is ${\bf C}^{2k}$:
the parameters are the centers of the solitons $z_1^{(i)}$ and the
modulus and phase of $c_i$, determining the scale and phase of each=
 individual
soliton.

Solutions of type B are given by the complementary ansatz:
\[ \begin{array}{ccc} \phi_1=3DA_1=3DA_2=3D0 & ,& \phi_2(x)=3D\phi_2(x_3,x_4=
) \\
    A_3(x)=3DA_3(x_3,x_4)  &; & A_4(x)=3DA_4(x_3,x_4) \end{array} \]
We now meet the system
\[ \begin{array}{lr}
 F_{34}(x_3,x_4)=3D|\phi_2(x_3,x_4)|^2 &    \\
 (D_3-iD_4)\phi_2(x_3,x_4)=3D0 & \hspace{6.4cm}(9) \end{array} \]
again leading to the Liouville equations with the right sign to obtain
non-singular
finite energy density solutions:
\[ \phi_2^{[k]}({\bar{z}}_2)=3D\frac{2 f'({\bar{z}}_2)V^2 ({\bar{z}}_2)}
{|V({\bar{z}}_2)|^2+|f({\bar{z}}_2)V^2 ({\bar{z}}_2)|^2},\hspace{1cm}
{\bar{z}}_2=3Dx_3-ix_4 .\hspace{1.9cm}  (10) \]
The euclidean action is
\[  S[\phi_2^{[k]}]=3D\frac{2\pi=
 k}{g^2}\lim_{L\to\infty}\int_{-\frac{L}{2}}^{
\frac{L}{2}}dx_1 \,\int_{-\frac{L}{2}}^{\frac{L}{2}}dx_2 ,\]
there is no need to change the sign of euclidean time, but the 'euclidean'
magnetic flux
\[ \Phi_M[\phi_2^{[k]}]=3D\int\int dx_3\, dx_4\, F_{34} =3D-\frac{2\pi}{g} k=
 \]
is negative, fitting the sign of the $m$-spin projection. The moduli space=
 of
these solutions is also ${\bf C}^{2k}$.

We can understand the topological origin of these solutions in the following
way: consider ${\bf R}^4$ as ${\bf R}^2\otimes {\bf R}^2$. The finite energy
density conditions, alternatively in each ${\bf R}^2$ according to the type=
 of
solutions, plus the conformal invariance of the equations make the problem
tantamount to solving the system on $S^2\times S^2$ \cite{gp-br}. We have
two topological numbers: $n_{12}=3Dk$, electric flux, and $n_{34}=3D-k$,=
 euclidean
magnetic flux, labelling the topological sectors of the theory. In fact we
have more: setting the $x_4$-coordinate to be euclidean time there are two=
 other
possibilities for splitting ${\bf R}^4$ as ${\bf R}^2\otimes {\bf R}^2$.
Choosing a basis in which $\sigma_1$ is diagonal and the K\"{a}hler form is
$\omega=3Ddx_2\wedge dx_3+dx_4\wedge dx_1$, similar ansatzes yield
\[ \begin{array}{ccl} F_{23}(x_2,x_3)=3D-|\phi_1(x_2,x_3)|^2 &
       (D_2+iD_3)\phi_1(x_2,x_3)=3D0 & \hspace{.5cm}:\,\mbox{type A} \\
      F_{14}(x_2,x_4)=3D|\phi_2(x_1,x_4)|^2 &
       (D_1-iD_4)\phi_2(x_1,x_4)=3D0 & \hspace{.5cm}:\,\mbox{type B} \\
       \end{array} \]
or $\sigma_2$ diagonal, $\omega_2=3Ddx_3\wedge dx_1+dx_2\wedge dx_4$,
\[ \begin{array}{ccl} F_{13}(x_1,x_3)=3D-|\phi_1(x_1,x_3)|^2 &
       (D_1+iD_3)\phi_1(x_1,x_3)=3D0 & \hspace{.5cm}:\,\mbox{type A} \\
      F_{24}(x_2,x_4)=3D|\phi_2(x_2,x_4)|^2 &
       (D_2-iD_4)\phi_2(x_2,x_4)=3D0 & \hspace{.5cm}:\,\mbox{type B} \\
       \end{array} \]

There are electrically charged solitons in 3 different planes, labelled by,
\[ e_i=3D\varepsilon_{ijk} n_{jk} \;\; ,\;\; n_{jk}=3D-n_{kj} \;\; ,\; \;=
 e_i \in
     {\bf Z}^+ \]
and also magnetically charged solitons, labelled by,
\[ m_i=3Dn_{i4} \;\; , \;\; m_i \in {\bf Z}^- \]
although the a 'priori' conserved topological numbers $m_i$ will not survive
the Wick rotation to real time.

We can extend the theory by modifying the spinor tensor/spinor tensor
interaction in the following way:
\[j_{\mu\nu}^{v_a}(x)=3D \frac{1}{2} \Psi_+^\dagger (x)\Sigma_{\mu\nu}\Psi_+=
 (x)
-\frac{1}{2} v_a^2 n_a^\dagger \Sigma_{\mu\nu}n_a, \]
no summation in $a$, with $n_1=3D\left( \begin{array}{c} 0\\0\\1\\0=
 \end{array}
\right)\; ,\;n_2=3D\left( \begin{array}{c} 0\\0\\0\\1 \end{array} \right)=
 \,$
and $a=3D1,2$. With a re-scale of variables $\Psi_+ \to v_a\Psi_+$, $A_\mu=
 \to
v_aA_\mu\,$ and $x_\mu \to \frac{1}{gv_a}x_\mu$, the ansatz of type A
plus $v_2=3D0$, produces the system:
\[ \begin{array}{c} F_{12}=3D1-|\phi_1|^2 \\  (D_1+iD_2)\phi_1=3D0=
 \end{array}
\hspace{6.2cm} (11) \]
Solutions of (11) such that $\lim\limits_{r_1\to\infty}\phi_1(x_1,x_2)=3D1$
exist and have been thoroughly analysed in \cite{tau}. They are electric
vortices, flux tubes, with the flux concentrated around the zeroes of
$\phi_1(x)$: there is neither freedom of scale (because the theory is not
conformally invariant) nor phase freedom (because $U(1)_d$ symmetry
is broken spontaneously). The moduli space is ${\bf C}^k$ and the same
$e_i=3D\varepsilon_{ijk}n_{jk}$ topological quantum numbers are conserved,
$e_i\in {\bf Z}^+$.

Alternatively, the ansatz of type B and $v_1=3D0$ leads to the system:
\[ \begin{array}{c} F_{34}=3D|\phi_2(x_3,x_4)|^2-1 \\ =
 (D_3-iD_4)\phi_2(x_3,x_4)
=3D0 \end{array}
\hspace{5.5cm} (12) \]
and there are 'euclidean' magnetic anti-vortices of negative integer flux
concentrated around the zeroes of $\phi_2$. The moduli space is ${\bf C}^k$
and the topological quantum numbers $m_i=3Dn_{4i}\in {\bf Z}^-$ will not be
conserved because of the euclidean time character of $x_4$.

\section{}
In order to study the phase structure of the model, it is convenient to have=
 a
richer set of solutions. Adding a left handed Weyl spinor $\Psi_-$, the=
 anti-
monopole field with $m$-spin projection $-1/2$, a system of equations like
(3.a)-
(3.b) with $F^+$ replaced by $F^-$ arises. There are solutions of the kind
(8) and (10) with electric and magnetic fluxes of opposite sign.

Similar solutions to those described above have been studied by the author
in Reference \cite{gui}. In that case, the Hopf invariant forced flux tube
pairs.
 Also 't Hooft electric and magnetic flux lines found in pure Yang-Mills in=
 a
box \cite{hft} share many physical features with the Seiberg-Witten=
 solutions.
We study the quantum behaviour of the last type of solitons in 't Hooft's
framework.

 In the $A_4=3D0$ gauge,consider the operator
\[ \hat{B}(C,t)=3D\mbox{exp}[ig(\int d^3x[\hat{\vec{E}}(\vec{x},t)\int_C=
 ds\,
{\vec{A}}^A (\vec{y} (s))+{\hat{\Pi}}_1^* (\vec{x},t)\int_C ds\, \phi^A_1
 (\vec{y}(s))])] \]

\[ {\hat{\Pi}}_1 (\vec{x},t)=3D-i({\dot{\hat{\phi}}}_1(\vec{x},t)+\frac{1}{2=
}
 \hat{D_{z_1}\phi_2}(\vec{x},t)+\hat{D_3\phi_1}(\vec{x},t)), \]
which creates an electric flux smeared by $\phi_1$ along the curve $C$: a=
 type
A solution for which the zero of the $\phi_1$ field with multiplicity $e_3$
is repeated along $C$. Choosing $C$ as the $x_3$-axis, perhaps with periodic
boundary conditions $x_3(-L_3)=3Dx_3(L_3),\,\hat{B}(C,t)$  creates an
electric solitonic string of electric flux $n_{12}=3De_3$ which is conserved
for topological reasons. On coherent states of the Hilbert state; the action
of $\hat{B} (C,t)$ is
\[ \hat{B} (C,t)|\vec{A},\vec{E};\phi_1,\phi_2 {\rangle}_0 \propto |\vec{A}+
L_3A^A(x_1,x_2),\vec{E}; \phi_1+L_3\phi^A_1(x_1,x_2),\phi_2 {\rangle}_{e_3}=
 \]
The proportionality factor is due to the fact that $\hat{B}(C,t)$ measures
magnetic flux.

Quantum states related to type B solutions are more difficult to analyse
because they correspond to a tunnel effect between different vacua when the
euclidean time component $x_4$ is Wick-rotated to real time. In fact there
are non-homotopically trivial gauge transformations of the form:
\[ \Omega_{k_3} (\vec{x})=3De^{\frac{i2\pi ga_3}{L_3}x_3} \Omega(x_1,x_2),
\;\;\; ga_3=3Dk_3 \]
These act on the quantum states by means of an unitary operator:
\[ {\hat{\Omega}}_{k_3}(\vec{x}) | \vec{A},\vec{E};\phi_1,\phi_2=
 {\rangle}_{e_3}
 =3De^{i\omega (k_3)}|\vec{A},\vec{E};\phi_1,\phi_2 {\rangle}_{e_3} \]
 The group law ${\hat{\Omega}}_{k_3} {\hat{\Omega}}_{k_3^\prime}=3D
 {\hat{\Omega}}_{k_3+k_3^\prime}$, $\omega(k_3)+\omega(k_3^\prime)=3D
 \omega(k_3+k_3^\prime)$ and the requirement of the same action for all the
gauge
 homotopy classes $\omega(k_3)=3D\omega(k_3+n)$ yields: $\omega(k_3)=3D2\pi
 \theta_3 k_3$. To understand the physical origin of this angle , consider=
 the
 'smeared' Wilson operator:
\[ \hat{A}(C,t)=3D\mbox{exp}[ig(\int d^3x[\hat{\vec{A}}(\vec{x},t)\int_C=
 ds\,
{\vec{E}}^B (\vec{y} (s))+{\hat{\Pi}}_2^* (\vec{x},t)\int_C ds\, \phi^B_2
 (\vec{y}(s))])] \]

\[ {\hat{\Pi}}_2 (\vec{x},t)=3D-i({\dot{\hat{\phi}}}_2(\vec{x},t)+\frac{1}{2=
}
 \hat{D_{{\bar{z}}_1}\phi_1}(\vec{x},t)+\hat{D_3\phi_2}(\vec{x},t)), \]
$\hat{A}(C,t)$ creates a solution of type B, in the $A_4$-gauge, along $C$.
For $T$ long enough and $C\equiv x_3$-axis,
\[ \hat{A} (C,t)|\vec{A},\vec{E};\phi_1,\phi_2 {\rangle}_{e_3} \propto
|\vec{A}+ \Omega(\vec{x},T),\vec{E}; \phi_1\Omega(\vec{x},T),\phi_2
{\rangle}_{e_3} \]
\[ \lim_{t\to {-\frac{T}{2}}}\Omega(\vec{x},t)=3D\Omega_0(\vec{x}),\;
 \lim_{t\to {\frac{T}{2}}}\Omega(\vec{x},t)=3D\Omega_{k_3}(\vec{x}),\;=
 k_3=3Dm_3 \]
$\hat{A}(C,t)$ creates magnetic flux but the tunnel effect, and the=
 associated
instanton angle,
means that only strips with flux parametrized by $\theta_3$ are conserved.

Due to the other possible choices of splitting ${\bf R}^4$ in ${\bf R}^2
\times {\bf R}^2$ there are quantum states characterized by $(\vec{e},
\vec{\theta})$, 3 integers and 3 angles, corresponding to the solutions of=
 the
Seiberg-Witten solutions. This analysis is a semi-classical one but=
 sufficient
because it is the regime within which we trust the theory. The free energy
of each
state is thus known and the phases are as follows:
\begin{enumerate}
\item $v_a^2=3D0$. Both electric and magnetic flux spread out over all the=
 space
and we are in the Coulomb phase.
\item $v_1^2=3D0$. In this phase there is a constant background=
 $\vec{\theta}$
magnetic field
in the $\vec{\theta}$-vacua ground states. In the $SPA$,
\[ \langle \theta_3| e^{-{\hat{H}}T}|\theta_3\rangle \propto
\mbox{exp} \{ e^{-2\pi L_1L_2/g^2} KL_3T\cos \theta_3\} \]
where $K$ is a determinantal factor accounting for the gauge and spinor
fluctuations in the presence of an instanton up to the one-loop order.
\item $v_2^2=3D0$. There is electric order: electric vortices of integral
magnetic charge exist and we
are in a phase of electric charge confinement.
Note that due to the peculiar $m$-spin properties of the solitons only
massless particles are
confined.
\item Finally, if $v_a^2\le 0$, there are no solutions of the first order
equations. This is the normal phase.

{\bf ACKNOWLEDGEMENTS}

The author has greatly benefited from conversations with S.Donaldson and
O.Garc\'\i a-
Prada on the mathematics of the Seiberg-Witten invariants. Financial support
by the DGICYT under contract PB92-0308 is acknowledged.
\end{enumerate}

\end{document}